\documentclass[conference]{IEEEtran}
\usepackage{amsmath,epsfig}
\usepackage {graphicx,amssymb,rotating,cite}
\usepackage{cite,epsfig,calc,ams}
\newtheorem{Proof}{Proof}
\newtheorem{thm}{Theorem}

\begin{document}

\title{Full MIMO Channel Estimation Using A Simple Adaptive Partial Feedback Method}

\author{\IEEEauthorblockN{Kamal Shahtalebi}
\IEEEauthorblockA{Department of Information Technology\\
The University of Isfahan, Isfahan\\ Iran, Postal Code:
81746-73441 \\
Email: shahtalebi@eng.ui.ac.ir} \and \IEEEauthorblockN{Gholam Reza
Bakhshi} \IEEEauthorblockA{Department of Electrical and Computer\\
Engineering, Yazd University, Yazd, Iran\\
Email: farbakhshi@yahoo.com} \and \IEEEauthorblockN{Hamidreza
Saligheh Rad}
\IEEEauthorblockA{School of Engineering and Applied\\
Sciences, Harvard University\\ Cambridge MA, 02138, USA\\
Email: hamid@seas.harvard.edu}}

\maketitle

\begin{abstract}

Partial feedback in multiple-input multiple-output (MIMO)
communication systems provides tremendous capacity gain and enables
the transmitter to exploit channel condition and to eliminate
channel interference. In the case of severely limited feedback,
constructing a quantized partial feedback is an important issue. To
reduce the computational complexity of the feedback system, in this
paper we introduce an adaptive partial method in which at the
transmitter, an easy to implement least square adaptive algorithm is
engaged to compute the channel state information. In this scheme at
the receiver, the time varying step-size is replied to the
transmitter via a reliable feedback channel. The transmitter
iteratively employs this feedback information to estimate the
channel weights. This method is independent of the employed
space-time coding schemes and gives all channel components.
Simulation examples are given to evaluate the performance of the
proposed method.

\end{abstract}

\section{Introduction}

Multiple-input multiple-output (MIMO) communication systems have
recently drawn considerable attention in the area of wireless
communications as they promise huge capacity increase and bandwidth
efficiency. In the derivation of MIMO fading channel capacity, it is
usually assumed that the channel state information (CSI) is known at
the receiver,\footnote{The CSI can be estimated at the receiver by
transmitting a training sequence from the transmitter.} but not at
the transmitter, and the transmitter has no knowledge about the
channel coefficients. In fact, channel capacity can be tremendously
increased by adding antennas and/or perfect channel estimation at
the transmitter. Although increasing the number of antennas improves
the capacity of the system, but it noticeably adds to the
complexity, the transmitted power and the design and the production
expenses at the transmitter. Instead of adding more number of
antennas, one can realize that even a very limited amount of
information about the channel enables the transmitter to estimate
the channel and to improve the system capacity and the output bit
error-rate. The channel information can be fed back to the
transmitter via a reliable feedback channel. In the literature,
partial feedback refers to a simple feedback whereby simple, but
important information of the channel is sent from the receiver to
the transmitter. In some of the reported works, partial feedback
contains a real phase parameter that is a function of all channel
gains. This partial information allows the transmitter to employ
orthogonal space-time block codes (OSTBC) \cite{MGD2006}. In some
others, partial information is employed for some forms of antenna
selection. The antenna selection schemes optimally choose a subset
of available transmit and/or receive antennas according to some
selection criterion, and then process the signals associated with
the selected antennas \cite{GNP2000}. Hybrid methods are engaged in
some of other reported works in the related literature
\cite{GG2003,GGP2003}. Almost in all forms of closed-loop transmit
diversity schemes, i.e. communication systems equipped with partial
feedback loop, perfect CSI is not achieved and the transmitter only
has partial information about the channel. In such cases,
transmission structures based on known CSI, e.g. optimal beamforming
scheme (OBS), can not be used. To have a perfect CSI estimation, in
this paper we propose a simple feedback scheme in which the time
varying step-size of a proposed least square adaptive algorithm is
sent back to the transmitter. The transmitter exploits it to
recursively estimate the CSI. Therefore, having a satisfactory
estimate of CSI anables the transmitter to employ OBS schemes.

The proposed method is made and presented for multiple-input
single-output (MISO) systems. However, the generalization of the
algorithm for MIMO systems is easily possible. The rest of this
paper is organized as follows: In section~\ref{S-M}, the system
model is introduced. Simulation examples are given in
Section~\ref{S-M}. Finally, the paper is concluded in
Section~\ref{Conc}.

\section{System Model}\label{S-M}

The following theorem has an essential role in our proposed method.

\begin{thm}\label{Th1}
Assume the $n_T\times 1$ vector $H$ and the sequence
$\{X_k\}_{k=1}^N$ are known. Define
\begin{equation}
\label{e1} H_k=H_{k-1}+\mu_k X_k, ~~~~~~k=1,2,\cdots,
\end{equation}
where $H_0$ is an arbitrary vector. Then, for all real $\mu_k$
with\footnote{When $2\mbox{Re}(X^H_k(H-H_{k-1}))<0$, we must
exchange its position with $0$ in relation (\ref{em}).}
\begin{equation}
\label{em} \mu_k\in(0,2\mbox{Re}(X^H_k(H-H_{k-1}))/\|X_k\|^2),
\end{equation}
$\|H-H_k\|$ is a decreasing sequence. Hence, the optimum value of
$\mu_k$ to minimizes it is given by
\begin{equation}
\label{e2} \mu_k=\frac{\mbox{Re}(X^H_k(H-H_{k-1}))}{\|X_k\|^2}.
\end{equation}
In this case we have
\begin{equation}
\label{e3}
\|H-H_k\|^2=\|H-H_{k-1}\|^2-\frac{\mbox{Re}^2(X^H_k(H-H_{k-1}))}{\|X_k\|^2}.
\end{equation}
\end{thm}
\begin{Proof}See appendix \ref{Appendix I}.
\end{Proof}

In a MISO system with $n_T$ transmit antennas and $1$ receive
antenna, the $n_T\times 1$ channel vector $H$ describes the CSI. The
$i^{\mbox{\small th}}$ component of $H$ which is denoted by $h_{i}$
represents the channel fading coefficient from the $i^{\mbox{\small
th}}$ transmit antenna to the receive antenna. In many real systems,
the CSI (or $H$) is not fully provided to the transmitter. This is
due to some basic limitations like limited capacity of feedback
channel or rapid channel variations. Therefore, if the communication
system is based on known CSI assumption at the transmitter, its
estimated value or $\hat{H}$ is alternatively used. Assume
$\|H-\hat{H}\|\le\zeta$, where $\|.\|$ denotes the Euclidean vector
norm, $\zeta$ is a threshold level and its exact value depends on
the system characteristics. Now assume that holding
$\|H-\hat{H}\|\le\zeta$, the overall system has a satisfactory
performance. Now suppose $\|H-\hat{H}\|>\zeta$ which may lead to a
bad system performance. The proposed method in this paper starts to
come into the role at this moment. It reduces the difference between
$H$ and $\hat{H}$ till the point that the corresponding difference
comes under $\zeta$. The details of the proposed method are given
based on the Theorem~\ref{Th1}.

When $\|H-\hat{H}\|>\zeta$, the receiver sends a signal to the
transmitter indicating that the transmitter should stop data
transmission and both parties run the procedure (\ref{e1}) in which:
\begin{itemize}
\item $\{X_k\}_{k=1}^N$, a pseudo white $n_T\times 1$ training vector sequence,
is known at both transmitter and receiver.
\item the initial value is $H_0=\hat{H}$.
\item The step-size $\mu_k$ is the quantized value of (\ref{e2}). Note that
the right hand side of (\ref{e2}) is computed only at the receiver
and its quantized value replied to the transmitter via a partial
feedback channel. Our suggestion is the $3$ bits Lloyd quantization
method \cite{Llyod} that we have used in our simulations to send the
quantized value of $\mu_k$.
\end{itemize}

If for $k=n$, $H_n$ satisfies $\|H-H_n\|\le\zeta$, the transmitter
chooses $H_n$ as its new value for the channel estimate $\hat{H}$.
This selection happens by receiving the \emph{procedure ending
signal} from the receiver and then, the communication system will go
back to its normal operation. Note that when $\{X_k\}_{k=1}^N$ is a
pseudo white sequence (as it is), in almost all iterations we have
$\mbox{Real}\{X^H_k(H-H_k)\}\ne 0$ (see the recursion (\ref{e3})).
Therefore, $\|H-H_k\|^2$ is a decreasing sequence and for
sufficiently large $N$, we have
\begin{equation}
\label{e4} \lim_{k\rightarrow N} \|H-H_k\|\approx0.
\end{equation}

\begin{itemize}
\item The proposed method is independent of system coding
and modulation schemes. Hence, it can be useful in all MISO systems
equipped with partial feedback.
\item Generalization of the proposed method for MIMO systems with $n_R$
receiver antennas ($n_R>1$) is done by sending $n_R$ step-size
values back to the transmitter. In other words, there is one
step-size value for each corresponding receiver antenna.
\item Usually the feedback path is a narrow band channel.
The required bandwidth for feedback is proportional to the CSI
variations and the processing speed of both transmitter and
receiver.
\item In order to have a good performance, it is reasonable to have interruption in
symbol transmission by the transmitter during the channel estimation
phase. However, as recursion (\ref{e3}) shows, in each time instance
$k$, the transmitter takes an access to a better channel estimate
than that of the time instance $k-1$. Hence, if the transmitter
insists to continuously send its symbol and does not break it (even
when the channel estimate error $\{\|H-H_{k-1}\|\}$ is over the
threshold level $\zeta$), it is reasonable that at each time instant
$k$, the old channel estimate $H_{k-1}$ is replaced by its new one
$H_k$ at the transmitter. This is because the channel squared error
is a decreasing sequence.
\end{itemize}

\section{Simulations}\label{S-M}

\begin{figure} \centering
\shortstack[cc]{\includegraphics[width=.45\textwidth]{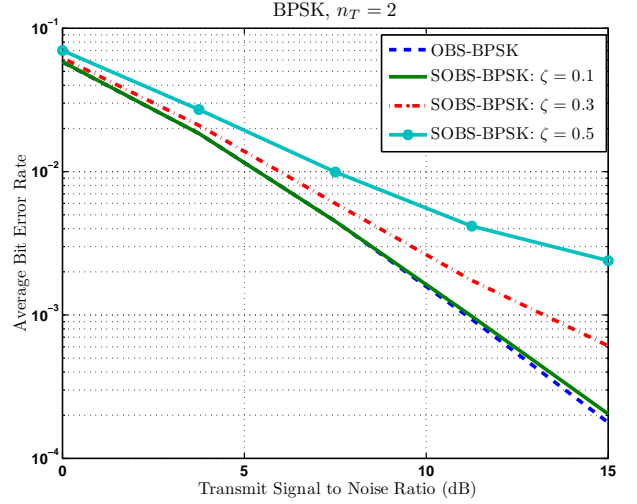}\\
(a) $n_T=2$} \shortstack[cc]{\includegraphics[width=.45\textwidth]{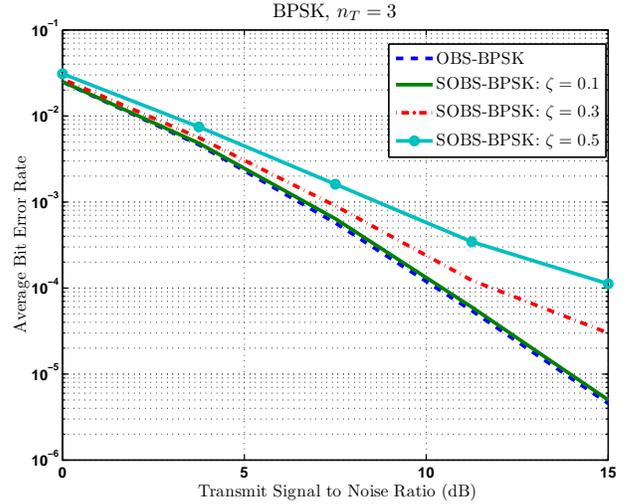}\\
(b) $n_T=3$} \caption{Comparison BER versus TNR for OBS and SOBS
systems, using BPSK modulation (a) $n_T=2$ (b) $n_T=3$ }
\label{Fig2}
\end{figure}

In our simulation we consider a MISO system with flat Rayleigh
fading channel in two cases. In case one, a binary phase shift
keying (BPSK) modulation and in case two, a quadrature phase shift
keying (QPSK) modulation are assumed. For each case we consider two
values $n_T=2$ and $n_T=3$ which led to four different situations.
Figures~\ref{Fig2} and ~\ref{Fig3} compare the bit error rate (BER)
versus the transmit power to noise ratio (TNR) of the OBS (see
Appendix~\ref{OBS}). Suboptimal beamforming scheme (SOBS) is used in
which the unknown $H$ is replaced by its estimated value $\hat{H}$
which has been given from the proposed method for $\zeta=0.1, 0.3,
0.5$.We assume no delay in feedback channel. Also it is assumed that
along the channel estimation procedure, the symbol transmission is
interrupted till the time the channel estimate error comes below
$\zeta$. Figure~\ref{Fig2} shows that with BPSK modulation and
$\zeta=0.1$, both OBS and SOBS have the same performance. As it is
expected and the histogram diagrams of the step-sizes for $n_T=2$
and $n_T=3$ (Figures~\ref{Fig4}a and ~\ref{Fig4}b) show, the
step-size curve has a normal shape around zero.

\section{Conclusion}\label{Conc}

In this paper, a full channel estimation procedure for MIMO systems
is proposed. In the suggested method, an iterative adaptive
algorithm is executed at both the receiver and the transmitter. The
required step-size value for the transmitter is sent back to the
transmitter by means of the feedback channel.

\appendices

\section{The optimum value of $\mu_k$}\label{Appendix I}

By subtracting $H$ from both sides of recursion (\ref{e1}), we have
\begin{equation}
\label{A1} H-H_k=H-H_{k-1}-\mu_k X_k.
\end{equation}
From equation (\ref{A1}), $\|H-H_k\|^2=(H-H_k)^H(H-H_k)$ is equal
to
\begin{equation}
\label{A2}
\|H-H_k\|^2=\|H-H_{k-1}\|^2+\mu_k^2\|X_k\|^2-2\mu_k\mbox{Real}\{(H-H_k)^HX_k\}.
\end{equation}
The left hand side of (\ref{A2}) is a decreasing sequence when
$\mu_k$ satisfies inequality (\ref{em}) and is minimized when
$\mu_k$ is given by (\ref{e2}). From equations (\ref{e2}) and
(\ref{A2}), we easily get (\ref{e3}).

\section{Optimal Beamforming Scheme}\label{OBS}

Assuming the known CSI, in optimal transmit beamforming scheme, the
$i^{\mbox{\small th}}$ transmitted signal is multiplied by the
normalized conjugate of the $i^{\mbox{\small th}}$ channel
coefficient. In other words, the beamformer vector
$\frac{1}{\|H\|}H^H$ is applied at the transmitter, where $(.)^H$
denotes the conjugate transpose. It means that if $s$ is the
transmit symbol, the received signal is
\begin{equation}
\label{e5} d=\frac{sH^HH}{\|H\|}+v,
\end{equation}
or equivalently
\begin{equation}
\label{e6} d=\sqrt{\sum_{i=1}^{N_T}h_i h^*_i} s+v,
\end{equation}
where $v$ is the additive Gaussian noise with zero mean and variance
$\sigma^2_v$.

\begin{figure} \centering
\shortstack[cc]{\includegraphics[width=.45\textwidth]{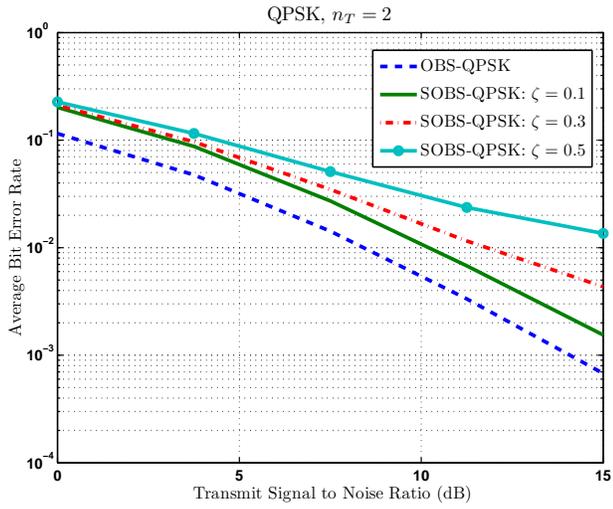}\\(a)
$n_T=2$}
\shortstack[cc]{\includegraphics[width=.45\textwidth]{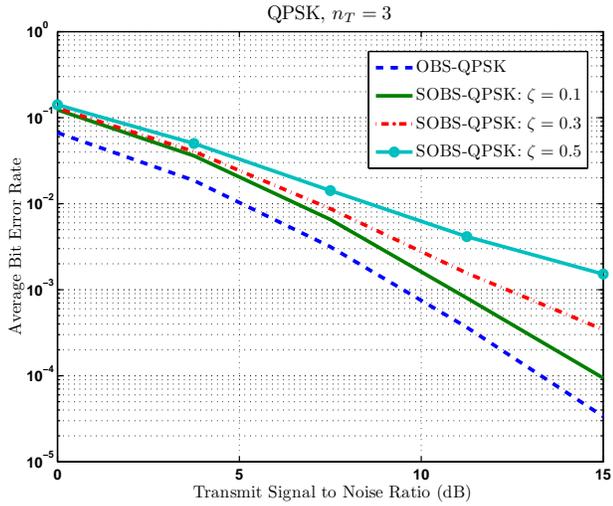}\\(b)
$n_T=3$}
\caption{Comparison BER versus TNR for OBS and SOBS systems,
using QPSK modulation with Gray code (a) $n_T=2$ (b)
$n_T=3$}\label{Fig3}
\end{figure}

\begin{figure} \centering
\shortstack[cc]{\includegraphics[width=.45\textwidth]{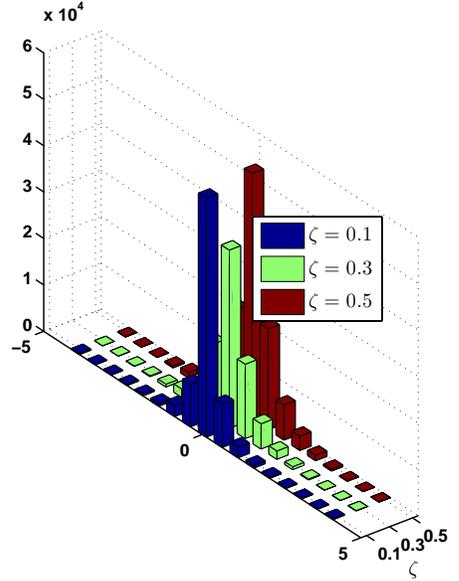}\\
(a) $n_T=2$} \shortstack[cc]{\includegraphics[width=.45\textwidth]{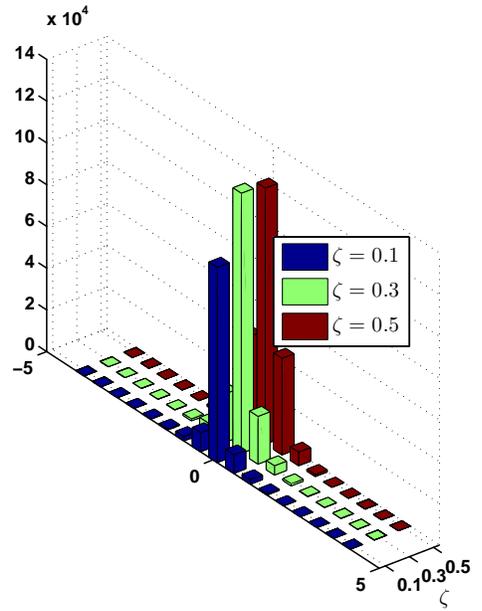}\\
(b) $n_T=3$} \caption{The histogram of the step size for (a) $n_T=2$
(b) $n_T=3$ } \label{Fig4}
\end{figure}


\begin{thebibliography}{8}
\bibitem{MGD2006}
J. K. Milleth, K. Giridhar, and D. Jalihal, ``On Channel
Orthogonalization Using Space-Time Block Coding With Partial
Feedback,'', {\em IEEE Trans.\ on Communications}, no.\ 6, pp.\
1121-1130, June 2006.
\bibitem{GNP2000}
D. Gore, R. Nabar, and A. Paulraj, ``Selecting an optimal set of
transmit antennas for a low rank matrix channel,'', in {\em
Acoustics, Speech, and Signal Processing, 2000. ICASSP'00.
Proceedings. 2000 IEEE International Conference on,} vol. 5,
(Istanbul), PP. 2785-2788, 2000.
\bibitem{GG2003}
M. \ Gharavi-Alkhansari and A.\ Greshman, ``Fast antenna selection
in MIMO systems,'' {\em IEEE Transactions on Signal Processing,}
vol. 52, no. 2, PP. 339-347, Feb. 2003.
\bibitem{GGP2003}
A. \ Gorokhov, D. A. \ Gore, and A. J. \ Paularj, ``Receive
antenna selection for MIMO spatial multiplexing: theory and
algorithms,'' {\em IEEE Transactions on Signal Processing,} vol.
51, PP. 2796-2807, Nov. 2003.
\bibitem{Llyod}
S. Lloyd, ``Least squares quantization in PCM,'' {\it IEEE
Transactions on Information Theory},vol. 28, no. 2, pp. 129-137,
1982.
\end{thebibliography}
\end{document}